\documentclass[aps,prl,a4paper,reprint,superscriptaddress,showpacs]{revtex4-1}
\usepackage{graphicx,graphics,color,epsfig}
\usepackage{hyperref}
\usepackage{amsmath}
\usepackage{amsfonts}
\usepackage{amssymb}
\usepackage{appendix}

\begin{document}

\title{Spin-Valley Beam Splitter in Graphene} 

\author{Yu Song}
\email{kwungyusung@gmail.com}
\affiliation{Microsystems and Terahertz Research Center, China Academy of Engineering Physics,
Chengdu 610200, P.R. China}
\affiliation{Institute of Electronic Engineering, China Academy of Engineering Physics,
Mianyang 621999, P.R. China}

\author{Lei Xie}
\affiliation{Microsystems and Terahertz Research Center, China Academy of Engineering Physics,
Chengdu 610200, P.R. China}
\affiliation{Institute of Electronic Engineering, China Academy of Engineering Physics,
Mianyang 621999, P.R. China}

\author{Zhi-Gui Shi}
\affiliation{Microsystems and Terahertz Research Center, China Academy of Engineering Physics,
Chengdu 610200, P.R. China}
\affiliation{Institute of Electronic Engineering, China Academy of Engineering Physics,
Mianyang 621999, P.R. China}

\author{Shun Li}
\affiliation{Microsystems and Terahertz Research Center, China Academy of Engineering Physics,
Chengdu 610200, P.R. China}
\affiliation{Institute of Electronic Engineering, China Academy of Engineering Physics,
Mianyang 621999, P.R. China}

\author{Jian Zhang}
\affiliation{Microsystems and Terahertz Research Center, China Academy of Engineering Physics,
Chengdu 610200, P.R. China}
\affiliation{Institute of Electronic Engineering, China Academy of Engineering Physics,
Mianyang 621999, P.R. China}

\begin{abstract}
The fourfold spin-valley degenerate degrees of freedom in bulk graphene can support rich physics and novel applications
associated with multicomponent quantum Hall effects
and linear conductance filtering.
In this work, we study how to break the spin-valley degeneracy of electron beams \emph{spatially}. 
We propose a spin-valley beam splitter
in a gated ferromagnetic/pristine/strained graphene structure. 
We demonstrate that, in a \emph{full} resonant tunneling regime for all spin-valley beam components, the formation of quasi-standing waves can lead 
four giant lateral Goos-H\"{a}nchen shifts as large as the transverse beam width, while the interplay of 
the two modulated regions can lead differences of resonant angles or energies for the four spin-valley flavors, manifesting an effective spin-valley beam splitting effect. The beam splitting effect is found to be controllable by the gating and strain.
\end{abstract}

\date{\today} 
\maketitle


The fourfold spin-valley degenerate degrees of freedom in bulk graphene can support rich physics and novel applications
associated with multicomponent quantum Hall effects \cite{young2012spin,yang2010hierarchy,islam2016scheme}
and linear conductance filtering \cite{grujic2014spin,wu2016full,soodchomshom2011strain,wang2016valley}.
For the spin-valley quantum Hall effects,
it is experimentally demonstrated that, the approximate SU(4) isospin
symmetry of Landau levels (LL) in graphene can be broken
by interactions such as strong Coulomb interaction, Zeeman effect, and lattice scale interactions,
manifesting as quantum Hall isospin ferromagnetic states \cite{young2012spin}.
Several other schemes have also been proposed theoretically, such as
breaking the fourfold degeneracy of the central LL by valley-scattering random potential, Zeeman interaction,
and electron-phonon coupling \cite{yang2010hierarchy};
or generating quantum spin-valley Hall effect 
via quantum pumping by adiabatically modulating a magnetic impurity and an electrostatic potential
in strained graphene \cite{islam2016scheme}.

For the spin-valley filtering effect,
it is demonstrated that, strain in a graphene barrier with carrier mass and spin-orbit coupling
can enforce opposite cyclotron trajectories
for the filtered valleys in a spin-valley dependent gap, demonstrating simultaneous filtering
of both valley and spin \cite{grujic2014spin}.
In addition, in strained graphene with Rashba spin-orbit coupling and magnetic barrier,
full valley- and spin-polarization currents can be accessed simultaneously,
due to the coexistence of valley and single spin band gaps \cite{wu2016full}.
Also, the interplay of modulation fields in a graphene ferromagnetic/strained (spin splitting barrier)/ferromagnetic
structure \cite{soodchomshom2011strain,wang2016valley} 
are shown as possible schemes. 

While almost all these works concern about %
conductances (quantum Hall or linear response),
almost nothing is known about how to break the spin-valley degeneracy of electron \emph{beams} in graphene.



\begin{figure}[b]
\centering  \includegraphics[width=\linewidth]{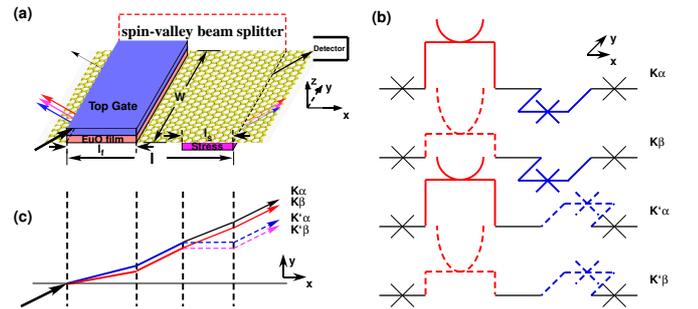}
\caption{Schematic diagrams of (a) the proposed spin-valley beam splitter,
(b) modulation profiles felt by electrons from different spin-valley flavours,
and (c) the motion paths in the device for electrons from different spin-valley flavours.
In (b) the modulations along the $z$ and $y$ axis stand for energy and wave vector, respectively.
}\label{fig2}
\end{figure}

In this work, we 
propose a spin-valley beam splitter
based on Goos-H\"{a}nchen (GH) shifts \cite{note1} and their difference between spin-valley flavours.
The device we considered is based on a proximity-induced ferromagnetic interaction
from a EuO film covered by a top gate \cite{haugen2008spin,yang2013proximity,song2015spin},
which lifts the spin degeneracy,
combined with a uniaxial strain, which lifts the valley degeneracy (see Fig. \ref{fig2}(a)).
We first consider the full quasi-ballistic transport regime.
It is shown that, the spin band-gaps and valley wave-vectors in the ferromagnetic and strained  regions
respectively mimic spin- and valley-dependent refractive indices,
acting as a spin or valley beam splitter near total reflections.
We subsequently consider the full quantum tunneling regime.
It is found that, resonant tunnelings can lead $\sim-\pi$ sudden phase jumps of the transmission components,
resulting four giant GH shifts as large as the transverse beam width \cite{note2}.
More importantly, the interplay of the spin- and valley-dependent imaginary wave vectors
in the ferromagnetic and strained regions lead to four different resonant angles (energies) for
each spin-valley flavours, thus demonstrating a spin-valley beam splitter
that lies in the intersection of spintronics and valleytronics.
The beam splitting behavior is found to be controllable by the gating as well as the strain.
Spin \emph{or} valley beam splitter in graphene devices was studied before \cite{zhai2011valley,zhang2015valley,wang2014spin,zhang2014spin};
however, the mechanism proposed in this work is novel, and previously unexplored.

Figure \ref{fig2}(a) shows the proposed device.
A graphene of length $L$ and width $W$ is grown on a substrate placed
in the $x$-$y$ plane.
An EuO film is deposited upon the graphene \cite{swartz2012integration} in the region of $(0,l_f)$,
with a top gate further grown on top of it.
A uniaxial strain is applied on the substrate in the region of $(l-l_{st},l)$.
$W$ is several times of $l$ to ensure that the edge effect is negligible \cite{tworzydlo2006sub}.
The spin- and valley-resolved trajectories of the reflection
and transmission beam components are shown in Fig. \ref{fig2}(a), with the line width approximately
standing for the beam strength.
For clearness, the sublattices position differences of $1/(k_F\cos\alpha)$ \cite{beenakker2009quantum}
are not shown ($k_F$ stands for the Fermi wave vector).
A detector placed at a proper position in the transmission region can be used to receive the splitting beam
component.

\begin{figure}[t]
\centering  \includegraphics[width=\linewidth]{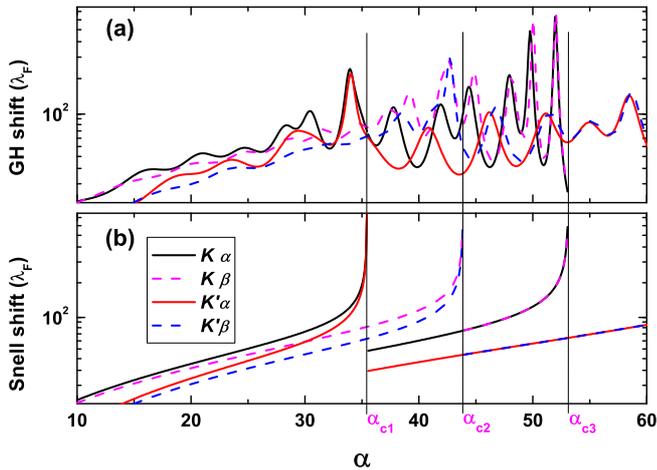}
\caption{(a) GH and (b) Snell shifts (in units of Fermi wave length) of the four beam components
as a function of incident angle at beam energy of $E$=20.
The parameters of the spin-valley beam splitter are $(l_f,l_w,l_{st})=(1,2,1), V_g=135.6$, and $A_{Sy}=4$.
}\label{fig7}
\end{figure}

An electron beam can be represented
as a wave packet of a weighted superposition of plane-wave
spinors \cite{li2003negative,beenakker2009quantum,song2013ballistic},
$\boldsymbol{\Psi}_{i}(x,y)=\frac{1}{\sqrt{2}}\int dq f(q-\bar{q})\Phi_p^+(x,y),$
$\boldsymbol{\Psi}_{r}=\frac{1}{\sqrt{2}}\int dq r_{\xi s}(q)f(q-\bar{q})\Phi_p^-(x,y),$
$\boldsymbol{\Psi}_{t}=\frac{1}{\sqrt{2}}\int dq t_{\xi s}(q)f(q-\bar{q})\Phi_p^+(x,y).$
Here the subscript $i,r,t$ stands for incident, reflection, and transmission, respectively,
$\bar{q}$ is the central transverse wave vector,
and $f()$ is the spectral distribution of the transverse wave vector,
which can be assumed to be of Gaussian profile.
For reflection and transmission, the weights
are further modulated by reflection/transmission coefficients that
are spin (with index $s=\pm1$ for spin up/down ($\alpha$/$\beta$))
and valley (with index $\xi =\pm1 $ for valley $K$ and $K^\prime$) dependent.
In the steady state approximation \cite{li2003negative,beenakker2009quantum,song2013ballistic},
the locus of the beam components can be given by
\begin{equation}
\sigma_{s\xi}^{r,t}=-\left(\frac{\partial \phi_{s\xi}^{r,t}}{\partial q_p}\right)_{E_p},\label{eq4}
\end{equation}
where $\phi_{s\xi}^{r}=\bar\phi_{s\xi}^{r}$ and $\phi_{s\xi}^{t}=\bar\phi_{s\xi}^{t}+k_p l$.
The phases relate with the coefficients through
$r_{\xi s}= |r_{\xi s}|e^{i\bar\phi^r_{\xi s}}$ and
$t_{\xi s}=|t_{\xi s}|e^{i\bar\phi^t_{\xi s}}$.

To obtain the phases, we calculate the transmission
coefficients by solving the right- and left-going
spinor eigenstates in the pristine ($\Phi_p^\pm$ shown in the wave packet),
ferromagnetic, and strained regions,
and subsequently using the well-know transfer matrix method \cite{born2000principles}
handling the continuity of them.
The eigenstates in each uniform region can be exactly resolved by decoupling
the two-order differential equation $H_j\Phi_j=E_j\Phi_j$.
The Hamiltonian in the pristine and strained ($j=p,s$) regions are well known \cite{katsnelson2006chiral,pereira2009strain}.
For the ferromagnetic ($j=f$) region, we adopt a half-metal model \cite{yang2013proximity,song2015spin,ang2016nonlocal}.
Rather than a simple Zeeman effect, this Hamiltonian roundly describes \cite{yang2013proximity,song2015spin}
the induced charge mass or opening energy gaps ($\Delta_s=(58+9s)$ meV),
re-normalized Fermi velocities ($v_s=(1.4825-0.1455s)v_F$),
and shifted Dirac points ($D_s=(-1.356+0.031 s)$ eV)
due to the proximity interaction,
that are all spin resolved.
The eigenstates in the pristine, ferromagnetic, and strained regions can be written in an uniform form
\begin{equation}
\Phi_j^\pm=e^{\pm ik_jx+iq_jy}
\left(
\begin{array}{cc}
\sqrt{E_j/2(\pm k_j+iq_j)}\\
\sqrt{(\pm k_j+iq_j)/2E_j}
\end{array}
\right),\label{eq2}
\end{equation}
where $+ (-)$ stands for the right- (left-) going propagation,
$E_p=E_{st}=E$ and $E_f=(E-\widetilde{D}_s+\Delta_s)/v_s$ with
$\widetilde{D}_s=D_s+V$ and $V$ the gate voltage.
In Eq. (2), $q_p=q_f=E\sin\alpha$ is the conserved transverse wave vector,
$q_{st}=q+\xi A_{Sy}$ with $A_{Sy}$ the $y$-component of the pseudo magnetic vector potentials \cite{pereira2009strain};
$k_p=\textmd{sign}(E_p)\sqrt{E_p^{2}-q_p^{2}}$,
$k_f=\textmd{sign}(E_f)\sqrt{E_f E_f^\prime-q_f^{2}}$ with $E_f^\prime=(E-\widetilde{D}_s-\Delta_s)/v_s$,
and $k_{st}=\textmd{sign}(E_{st})\sqrt{E_{st}^{2}-q_{st}^{2}}$.
For brevity, we express all quantities in dimensionless form by means of
a characteristic length $l_0 = 56.5$nm and energy unit $E_0 = 10$meV.

In Fig. \ref{fig7}, we look at the behavior of GH shifts in detail for the quasi-ballistic transport regime
(the beam energy is so high that all components are transported quasi-ballistically).
Also shown are the `electron paths' (see Fig. \ref{fig2}(c)) calculated in the geometric optics using Snell's Law.
The sine of refractive angles in the three regions read $\sin\alpha_p=q_p/E_p$,
$\sin\alpha_f=q_f/\sqrt{E_fE^\prime_f}$, 
and $\sin\alpha_{st}=q_{st}/E_{st}$.
Spin- or valley-dependent refractive indices can be defined as \cite{cheianov2007focusing,moghaddam2010graphene}
$n^f_s=\sqrt{(E_p-D_s-V)^2-\Delta_s^2}/(v_s|E_p|)$ and $n^{st}_\xi=q_p/(q_p+\xi A_{Sy})$.
Then the four Snell shifts are
$\sigma^{Snell}_{\xi s}=l_f\tan(\sin^{-1}(\sin\alpha/n_s))+l_w\tan\alpha+l_s\tan(\sin^{-1}(\sin\alpha/n_\xi))$.
It is clear that, the GH shifts simply oscillate around corresponding Snell shifts.
It is seen in Fig. \ref{fig7}(b) that, there are four critical angles corresponding to the four refractive indices,
i.e., $\sin\alpha_{c1}=n^f_\alpha$, $\sin\alpha_{c2}=n^f_\beta$, $\sin\alpha_{c3}=n^{st}_K$, and $\sin\alpha_{c4}=n^{st}_{K^\prime}$
($\alpha_{c4}$ is negative and not shown).
Near each critical angle, GH shifts 
of two beam components with a same spin or valley increases dramatically. 
This is an enhancement effect due to total reflection. 
As a result, spin beam splitter of $\alpha$/$\beta$ can be achieved near $\alpha_{c1}$/$\alpha_{c2}$;
while valley beam splitter of $K$/$K'$ near $\alpha_{c3}$/$\alpha_{c4}$.
In Fig. \ref{fig7}(a) it is also observed that,
the four Snell shifts are spin- and valley-dependent, 
especially for incident angle smaller than $\alpha_{c2}$. 
However, it is hard to achieve enough GH shift difference 
between any component and the other three ones.
Thus, spin-valley beam splitting is rather challenging in this transport regime.

\begin{figure}[t]
\centering  \includegraphics[width=\linewidth]{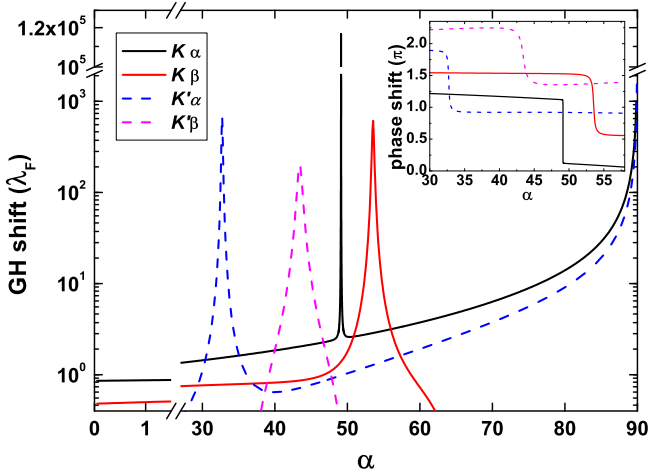}
\caption{The GH shifts (in units of Fermi wave length) of the four
beam components as a function of the incident angle at $E=2$.
The device parameters are the same as in Fig. \ref{fig7}.}\label{fig5}
\end{figure}

In Fig. \ref{fig5} we consider the behavior of GH shifts in the full resonant tunneling regime.
One can see that, as the incident angle increases, 
the two spin $\alpha$ (spin $\beta$) components undergo increasing
(decreasing) GH shifts approximatively proportional to
$\tan\alpha$ ($-\tan\alpha$).
This is still a quasi-ballistic transport behavior. 
It is distinct that, around some specific angles, the four GH shifts display sharp peaks.
The corresponding phases in the insert show sudden jumps of about $-\pi$.
This is a clear signal for a formation of a standing wave in resonant tunneling.
Recalling the condition for ideal standing waves in an infinite potential well, i.e.,
$\sin k_p l_w=0$, 
the resonant angles can be solved as $\alpha=38.2^\circ$ with an only allowed $n=1$.
The resonant angles for the four spin-valley flavours
($K\alpha$, $K\beta$, $K'\alpha$, and $K'\beta$) are 49.1$^\circ$, 53.5$^\circ$, 32.7$^\circ$, and 43.4$^\circ$, respectively.
They all deviate from the ideal resonant angle, because the standing waves are formed in
non-ideal potential wells formed by walls of spin band-gaps and valley wave-vectors, see Fig. \ref{fig2}(b).
In this case, electron waves can leak out through evanescent waves
in the walls.
This generally lowers the quantization energies and wave vectors, resulting larger resonant angles
(this is true except for the $K'\alpha$ case).
More importantly, 
interplay of the spin- and valley-dependent imaginary wave vectors
in the modulation regions (see Fig. \ref{fig2}(b)) lead different standing waves, 
implying different wave vectors and hence different resonant angles
for different spin-valley flavours.
At each resonant angle,
the resonant GH shift exceeds the transverse beam width 
and the other three ones are small.
When we tune the incident angle, the four resonant spin-valley beam components can be splitting in sequence,
manifesting a novel spin-valley beam splitting effect.


\begin{figure}[t]
  \centering
  \includegraphics[width=\linewidth]{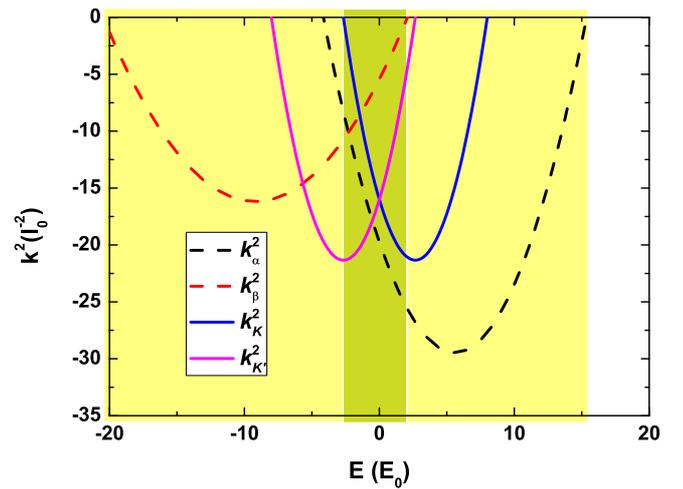}\\
  \caption{Square of the four wave vectors in the two modulated regions
  as a function of the beam energy.
  $\alpha$=30$^\circ$, $V_g=135.6$, and $A_{Sy}=4$.
}\label{fig4}
\end{figure}

It is surprising that, the spin band-gaps and valley wave-vectors play such different roles
for beam splitter in different transport regimes.
In Fig. \ref{fig4}, we plot the square of the four wave vectors as a function of beam energy for a fixed incident angle
(note that a beam component is related to one spin wave vector plus one valley wave vector).
Three energy ranges can be defined. 
i) The energy rang that exceeds 15 or -21. 
All wave vectors are real, 
meaning that all the four beam components transport through the sandwich structure quasi-ballistically.
This is the case we considered in Fig. \ref{fig7}, giving behaviors similar to geometric optics
and can be applied as spin or valley beam splitter near total reflection.
ii) The range between -2.7 and 2.1.
In this range, the four wave vectors all become imaginary,
implying that the four spin-valley beams all undergo evanescent transport or resonant tunneling.
This is the case we considered in Fig. \ref{fig5}, showing spin-valley beam splitter
due to flavours dependent standing waves.
This is the central result of the present work.
iii) The energy range between.
At least one wave vector (say $k_\beta$) is real.
The two related components ($K\beta$ and $K'\beta$) cannot undergo resonant tunneling.
As a result, they cannot be splitting from each other (although can be
splitting from the $K\alpha$ and $K'\alpha$ components).
In this case, only components $K\alpha$ and $K'\alpha$ can be splitting,
demonstrating a partial spin-valley beam splitting effect.

\begin{figure}[t]
\centering  \includegraphics[width=\linewidth]{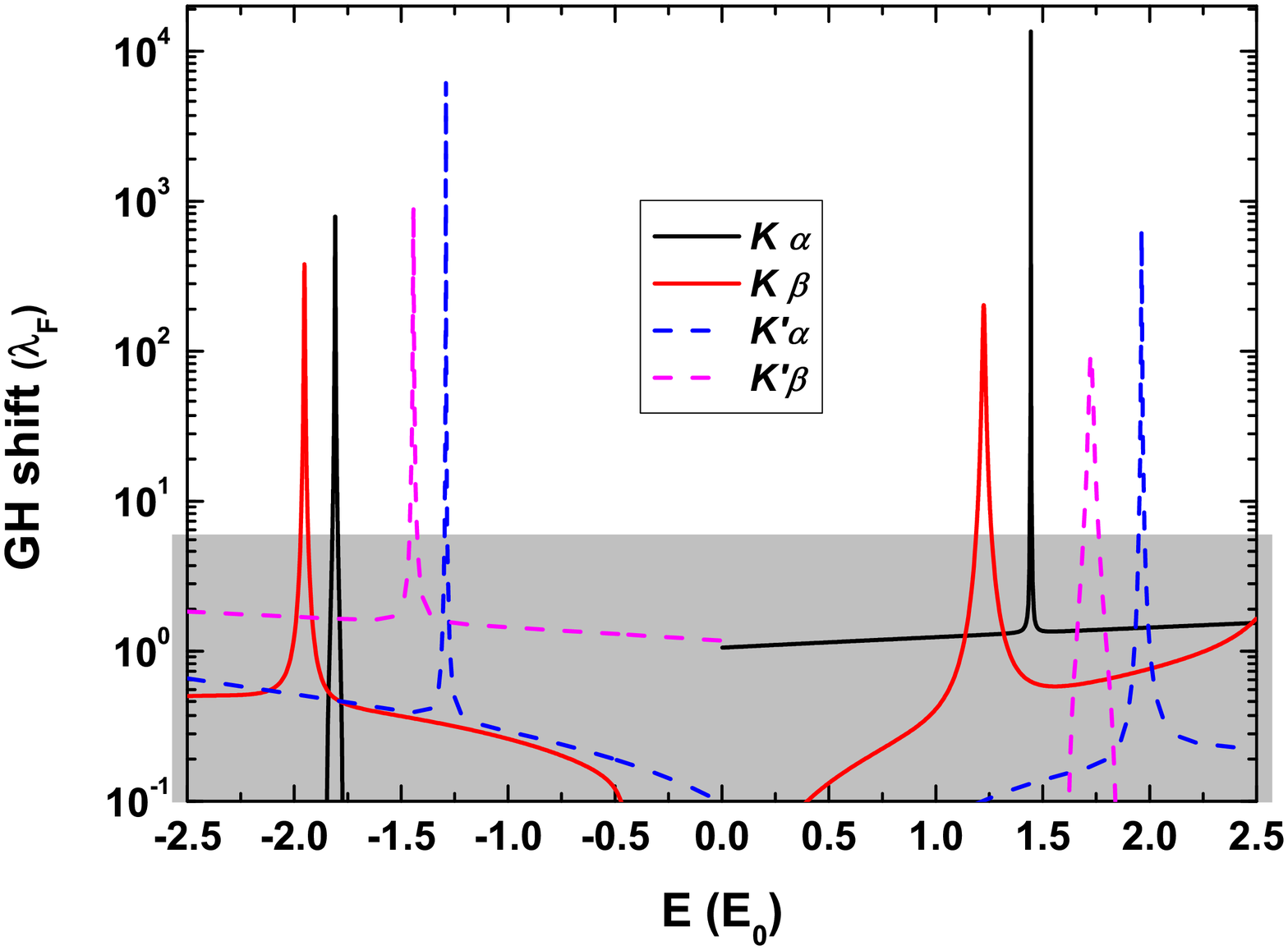}
\caption{The GH shifts (in units of Fermi wave length) of the four beam components as a function of the incident energy at $\alpha=30^\circ$.
The device parameters are the same as in Figs. \ref{fig7} and \ref{fig5}.
}\label{fig3}
\end{figure}

The full spin-valley beam splitting is further shown in the energy space in Figure \ref{fig3} .
One can see clear huge resonant peaks on background of GH shifts of the order of the Fermi wavelength.
The resonant energies are around the one of an ideal standing wave with $n=\pm1$,
i.e., $E_{resonant}=\pm1.81$.
The sequence of components is found to be in opposite orders
compared with that in the incident angle space (Fig. \ref{fig5}).
These behaviors mean that,
in addition to by tuning incident angle at fixed beam energy,
spin-valley beam splitting can also be achieved by tuning the beam energy at fixed incident angle.
This can have potential applications in splitting beam of wide energy distribution.

\begin{figure}[t]
\centering
\includegraphics[width=\linewidth]{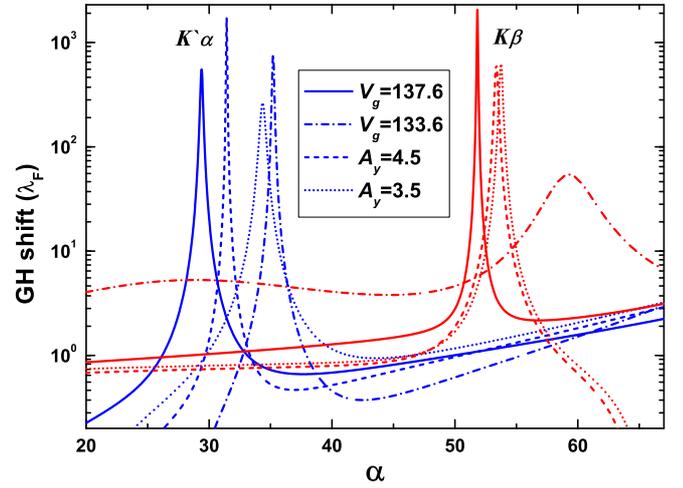}\\
\caption{GH shifts of the $K\beta$ and $K^\prime\alpha$ components
as a function of the incident angle at $E=2$ for
different gate voltages or strains. $l_f=1$, $l_w=2$, and $l_s=1$.
}\label{fig6}
\end{figure}

We at last consider the tunability of the full spin-valley beam splitter.
Figure \ref{fig6} shows the GH shifts of $K\beta$ and $K^\prime\alpha$ flavours
for different gate voltages or strains.
As can be seen, the resonant GH shift moves to a smaller incident angle
when a higher gate voltage or a bigger strain is applied.
This is because, the wave leaks become weaken and the energy of the quasi-standing
wave increases.
On the other hand, the resonant angle becomes bigger when a lower gate voltage or a smaller
stress is applied.
All these trends are the same for the other three components
(components $K\alpha$ and $K^\prime\beta$ are not shown for clearness).
The GH shifts of $\beta$-related flavours decrease at a lower gating
because the wave leaks increase.
The above results imply that,
the resonant angles of the splitting components can be chosen by tuning the gating or strain,
provided the transport remains in the full resonant tunneling regime. 
On the other hand, the bigger the strain, the wider the intersection set
(also limited by the two $K^f_s$ sets),
benefiting to the full resonant tunneling.

In summary, we have demonstrated that the spin-valley fourfold degeneracy of electron beam in
graphene can be totally eliminated by flavour-dependent giant GH shifts
in a full resonant tunneling regime. 
This manifests a novel application of spin-valley beam splitter, lying in
the intersection of spintronics and valleytronics.
The formation of standing waves in the potential wells, 
which leads sudden jump of phase of about $-\pi$,
and the interplay of the spin- and valley-dependent wave leaks, 
which leads difference in resonant angles,
are at the heart of the mechanism of the beam splitting effect.
The spin-valley beam splitting can be achieved by tuning incident angle
 at fixed beam energy or vice versa,
and can be controlled by gating or strain.

We have also shown that, the spin band-gaps and valley wave-vectors play
totally different roles in the full ballistic transport regime.
They act as medias with spin or valley dependent refractive index
and can be applied as spin or valley beam splitter near total reflections.

We encourage experimental works on the proposed device and mechanism.



This work was supported by the National Natural Science Foundation of
China (NSFC) under Grant No. 11404300
and the Science Challenge Project (SCP) under Grant No. JCKY2016212A503.



%

\end{document}